\documentclass{jfm}

\usepackage{graphicx}
\usepackage{natbib}
\usepackage{bm}

\ifCUPmtlplainloaded \else
  \checkfont{eurm10}
  \iffontfound
    \IfFileExists{upmath.sty}
      {\typeout{^^JFound AMS Euler Roman fonts on the system,
                   using the 'upmath' package.^^J}%
       \usepackage{upmath}}
      {\typeout{^^JFound AMS Euler Roman fonts on the system, but you
                   dont seem to have the}%
       \typeout{'upmath' package installed. JFM.cls can take advantage
                 of these fonts,^^Jif you use 'upmath' package.^^J}%
      }
  \else
  \fi
\fi

\ifCUPmtlplainloaded \else
  \checkfont{msam10}
  \iffontfound
    \IfFileExists{amssymb.sty}
      {\typeout{^^JFound AMS Symbol fonts on the system, using the
                'amssymb' package.^^J}%
       \usepackage{amssymb}%
       \let\le=\leqslant  
         
      }{}
  \fi
\fi

\ifCUPmtlplainloaded \else
  \IfFileExists{amsbsy.sty}
    {\typeout{^^JFound the 'amsbsy' package on the system, using it.^^J}%
     \usepackage{amsbsy}}
    {}
\fi

\newsavebox{\astrutbox}
\sbox{\astrutbox}{\rule[-5pt]{0pt}{20pt}}

\title[Flapping states of an elastically anchored wing]{Flapping states of an elastically anchored wing in a uniform flow}

\author[A. Orchini, A. Mazzino, J. Guerrero, R. Festa, C. Boragno]
{A.\ns O\ls R\ls C\ls H\ls I\ls N\ls I$^{2}$,\ns
A.\ns M\ls A\ls Z\ls Z\ls I\ls N\ls O$^{1,3}$\thanks{Email address for correspondence: andrea.mazzino@unige.it}, 
J.\ns G\ls U\ls E\ls R\ls R\ls E\ls R\ls O$^1$ \break
R.\ns F\ls E\ls S\ls T\ls A$^2$ \and
C.\ns B\ls O\ls R\ls A\ls G\ls N\ls O$^2$
}
\affiliation{$^1$
Dipartimento di Ingegneria delle Costruzioni, dell'Ambiente e del Territorio, 
Universit\`a di Genova, via Montallegro 1,  16145 Genova (Italy)\\[\affilskip]
$^2$Dipartimento di Fisica, Universit\`a di Genova,
via Dodecaneso 33, 16146 Genova (Italy) \\[\affilskip]
$^3$Istituto Nazionale di Fisica Nucleare, Sezione di Genova, 
via Dodecaneso 33, 16146 Genova (Italy)}

\pubyear{2010}
\volume{650}
\pagerange{119--126}
\date{?; revised ?; accepted ?. - To be entered by editorial office}
\begin{document}

\maketitle

\begin{abstract}
Linear stability analysis of an elastically anchored wing 
in a uniform flow
is investigated both analytically and numerically. The analytical formulation 
explicitly takes into account the effect of the wake on the wing
by means of Theodorsen's theory.
Three different parameters non-trivially rule the observed dynamics:
mass density ratio between wing and fluid, spring elastic constant 
and distance between the wing center of mass and the spring anchor 
point on the wing. We found relationships between these parameters 
which rule the transition between stable equilibrium and fluttering. 
The shape of the resulting marginal curve has been successfully 
verified by high Reynolds number direct numerical simulations. 
Our findings are of 
interest in applications related to energy harvesting by fluid-structure 
interaction, a problem which has recently attracted a great deal of 
attention. The main aim in that context is to identify the optimal 
physical/geometrical system configuration leading to large sustained motion,
which is the source of energy we aim to extract.
\end{abstract}

\begin{keywords}
Fluid-structure interaction, energy harvesting
%Authors should not enter keywords on the manuscript, as these must be chosen by
%the author during the online submission process and will then be added during
%the typesetting process (see
%http://journals.cambridge.org/data/\linebreak[3]relatedlink/jfm-\linebreak[3]keywords.pdf
%For the full list)
\end{keywords}

%%%%%%%%%%%%%%%%%%%%%%%%%%%%%%%%%%%%%%%%%%%%%%%%%%%%%%%%%%%%%%%%%%%%%
\section{Introduction}
\label{sec:0}
The study of the mutual interaction between fluids and elastic objects is
a problem of paramount importance in many fields of science
and technology \citep{dh_annual_01}.  In bio-fluid mechanics,
 with the advent of 
supercomputers,  it becomes a cornerstone for the quantitative understanding
of a variety of problems ranging from blood pressure interaction 
with arterial walls \citep{ams_06} and the blood interaction with 
mechanical heart valves \citep{verzicco09},
to animal 
locomotion and self-propulsion \citep{FL_06}
and aerodynamics of insect flights   \citep{sane_03}.\\ 
It is also a topic of growing interest 
in relation to the possibility of manipulating
the fluid flow to enhance 
aerodynamics performances of immersed bodies \citep{AB09}.\\
Fluid-structure interaction is also a crucial aspect in the design of 
many engineering systems, e.g., aircrafts and bridges. The failure in
 considering 
the effect of oscillatory interactions can be catastrophic. The ultimate
goal is thus to reduce at a minimum all sources of potentially resonant
couplings between fluid and structure.\\
If on the one hand the reduction of  the latter  mechanisms
is thus  a crucial need for the correct project of many engineering systems,
on the other hand 
situations exist where the enhancement of the same 
coupling mechanisms between the dynamics of flexible solids and 
surrounding air/water flows are strongly desired and sought
in order  to generate self-sustained, 
possibly large-amplitude,
motion of the solid body. This is a typical requirement for energy-harvesting
devices  through which solid body vibration can be successfully converted 
into electrical energy.\\
Among the many  strategies to achieve the goal of energy extraction 
from solid vibrations, the idea to harvest energy from the wind kinetic energy 
by flapping wings turned out to be  particularly attractive and fruitful
\citep{McL81}. It is beyond the scope of the present
paper to provide a review of 
developments which  followed from  the seminal paper by  \citet{McL81}.
There remains however  much work ahead of us, both on the side of material
science (to optimize material performances in term of elasticity and 
electrical efficiency) and on those of fluid mechanics and aeroelasticity.  
One of the main limitations of many of the
available ``pitch-and-plunge'' devices \citep{pitch}
is that 
one or more of the active degrees of freedom
of the system are explicitly driven (e.g., mechanically, by a motor) 
while the other available degrees of freedom are left free to interact
with the flow for the final aim of producing unceasing oscillations.
The external motor clearly reduces the net gain of the device.\\

Our aim here is to propose and study from the fluid mechanic point of view
a simple system composed by a rigid thin wing (actually, a rigid airfoil)
anchored to an elastic spring and free to move under the action of
a constant wind. By means of normal mode linear stability analysis
and direct numerical simulations (DNS)
we aim at investigating whether self-sustained oscillations
can be obtained for suitable choices of the 
available parameters (both physical and geometrical) 
and in the absence of any external motor.

The paper is organized as follow.
In Sec.~\ref{sec:1} we introduce our flapping system and the associated 
 equations of motion; in Sec.~\ref{linear} the linear stability analysis
is presented together with the obtained results;  in Sec.~\ref{dns}
we describe the numerical method we exploited both to verify the linear 
analysis predictions and to extend the study
to fully nonlinear regimes. The last section is reserved for conclusions.

%%%%%%%%%%%%%%%%%%%%%%%%%%%%%%%%%%%%%%%%%%%%%%%%%%%%%%%%%%%%%%%%%%%%%
\section{The flapping system}
\label{sec:1}
\subsection{Equations of motion}
\label{sec:1.1}
\begin{figure}
\centerline{\includegraphics[width=8cm,angle=-0]{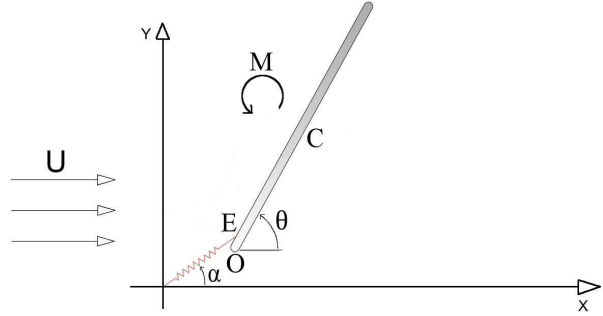}}
\caption{Cross-sectional sketch of the flapping device. $C$ denotes the wing center of mass, $O$ is the wing leading edge 
and $E$ is the anchor point of the elastic spring 
(supposed to have a zero rest length). The oriented curved 
arrows fix the 
convention on the positiveness of angles and moments. 
The angle $\alpha$ is magnified for the sake of clarity, the
analysis being mostly in the linear regime where the wing is almost aligned 
along the unperturbed flow.}
%The segment represents the  transverse size $L_y$.}
\label{fig1}
\end{figure}

The system we consider consists (see Fig.~\ref{fig1} for a  
cross-sectional view) 
of a rigid rectangular airfoil, of lenght (chord) $L$, 
thickness $\delta \ll L$, and span $S\gg L$. 
An elastic spring connects an external fixed point to an anchor point $E$
belonging to the longitudinal axis of the airfoil.
Such a structure is exposed to a uniform wind ${\bf U}$, 
blowing along the $ X$ axis (from left to right).\\
We will study the cross-sectional two dimensional motion, which approximately 
describes the fully three-dimensional motion of fluid and structure 
for an airfoil of very high aspect ratio $S/L$. 
We are interested in investigating the stability of the (trivial) equilibrium 
configuration corresponding to the airfoil aligned along the unperturbed 
wind field ${\bf U}$. Let us fix the origin of the $X$ axis so that 
the coordinate of the airfoil anchor point $E$ is in the origin at 
the equilibrium configuration ($X_E =0$). Indicating by $C$ the center of 
mass position, we define the ratio $r= \overline {CE}/L$, which will turn out 
an important parameter of the airfoil motion. Moreover, we will 
only consider situations corresponding to $E$ on the left (upwind) of $C$. 
A divergence instability is indeed trivially expected in the opposite case, 
i.e. when $E$ is on the right (downwind) of
$C$. In this work we assume that $C$ is in the wing midpoint 
(the whole analysis can be easily generalized to an arbitrary 
position of $C$), thus  $0 \le r \le 1/2$.

Let us now perturbate the equilibrium configuration and evaluate the 
elastic force ${\bf F}^e$ and its torque. If we conveniently choose
 $E$ as the pole for moments, then the elastic force does not
 produce any torque. Up to linear contributions in Taylor expansion,
%------------------------------------------------------------------------
%------------------------------------------------------------------------
the expression for ${\bf F}^e$ reads:
\begin{equation}
F^e_x=-k\, x_C 
\label{eq:felxl}
\end{equation}
\begin{equation}
F^e_y=-k\,(y_C-r\,L\, \theta) \;\;\; ,
\label{eq:felyl}
\end{equation}
where lowercase coordinates indicate perturbations with respect to the 
equilibrium configuration and $k$ is the spring constant.\\ 
The resulting equations of motion for the perturbations are:
\begin{eqnarray}
m\, \ddot{x}_C&=&-k\, x_C 
\label{eq:x}\\
m\, \ddot{y}_C&=&-k\, (y_C-r\,L\,\theta) - F^L
\label{eq:y}\\
%I_C \ddot{\theta}&=&  - m\,r\,L\, \ddot{y}_C + M^L_E
I_C \ddot{\theta}&=& -(r\, L)^2 \,k\,\theta + r\,L\,k\,y_C+r\,L\, F^L
+M^L_E
\label{eq:I}
\end{eqnarray}
where $I_C$ is the moment of inertia (around the center of mass axis), 
$m$ is the wing mass, $F^L$ and $M^L_E$ are the lift force and the lift 
momentum. 

\subsection{Aerodynamics forces and moments}
\label{sec:1.2}
The expression for $F^L$ and $M^L_E$ can 
be obtained by treating the point $E$ as one does for the flexural point 
in the classical `pitch-and-plunge' system \citep{pitch}
and then exploiting Theodorsen's theory
\citep{bah57}.  Because of the fact that 
the latter assumes a sinusoidal response of the
system (with no damping and no amplification), it is fully justified
to determine marginal curves in the parameter space separating stable regimes
from unstable ones (i.e.~the so-called flutter
condition). Determining such curves is one of the main aim of the
present paper.\\
According to Theodorsen's theory, the following expressions arise for  $F^L$
and $M^L_E$ \citep{bah57}:
\begin{equation}
% X_{L/4}   X_{L/2} X_{3 L/4}      
F_L= \rho\pi b^2 \left [\ddot{y}_E +  X_{L/2}\, 
\ddot{\theta}\right ]+\rho \pi b^2 U \dot{\theta}+
 \rho U^2L\pi
\left [\theta + \frac{\dot{y}_E}{U}+ X_{3 L/4} \,\frac{\dot{\theta}}{U}\right ]\,{\cal C}(q)
\label{lift}
\end{equation}
\begin{eqnarray}
M^L_E&=& - \rho U^2 L \pi  X_{L/4} \, 
\left [\theta + \frac{\dot{y}_E}{U}+ X_{3 L/4}    \frac{\dot{\theta}}{U}\right ]\,{\cal C}(q)
\nonumber \\
&-& \rho \pi b^2\, X_{L/2}\, \left [\ddot{y}_E+ X_{L/2}\, \ddot{\theta}\right ]-\frac{\rho \pi b^4}{8}\ddot{\theta}- X_{3 L/4}\,\rho\pi b^2 U\dot{\theta}
\label{moment}
\end{eqnarray}
where $b\equiv L/2$, and $X_{L/4}$,    $X_{L/2}$,  and $X_{3 L/4}$
are the distances of wing points at $L/4$, $L/2$ and $3 L/4$ 
from the leading edge.
All terms involving $b$ are the so-called added-mass terms.\\
The function ${\cal C}(q)$ is the Theodorsen's function 
through which the wake effect on the wing motion can be explicitly 
taken into account \citep{bah57}.
It depends on the
reduced frequency $q\equiv \omega\,L/(2 U)$, where $\omega$ is the
pulsation associated to the system harmonic response.
The form of ${\cal C}(q)$ can be given in terms of Hankel functions
\citep{GR} as:
\begin{equation}
{\cal C}(q) = \frac{H_1^{(2)} (q) }{H_1^{(2)} (q) + i\, H_0^{(2)} (q) }  \;\;\; .
\label{hankel}
\end{equation}
Eqs.~(\ref{eq:x})-(\ref{eq:I}) with the expressions for the aerodynamics 
force and moment (\ref{lift}) and (\ref{moment})
and the relationships  $x_C=x_E$, $y_C=y_E+r\,L\theta$ constitute
our equations of motion for $x_C$, $y_C$ and $\theta$
under the constraint of harmonic behavior, as
dictated by Theodorsen's theory.\\
%  This is not a limitation here because 
%we are just interested in identifying the fluttering condition
%separating stable from unstable regimes.\\
Some comments on the structure of the equations above are worth discussing.
The equation for $x_C$ is decoupled from the others; 
this variable can be thus ignored.
%, consistently
%with the initial assumption that $E$ is treated in the same ways as the
%the flexural point.
With respect to the classical `pitch-and-plunge' problem \citep{pitch}
here  $y_C$ and $\theta$ are two-way coupled in a stronger way,
a consequence of the elastic force acting on the wing
which jointly depends on both  $y_C$ and $\theta$.  This  is expected 
to cause interesting and peculiar behaviors.
\section{Linear stability analysis}
\label{linear}
Let us now investigate the normal mode linear stability analysis of the system.
This can be easily done by normal mode decomposition:
\[ y_C(t)=Y\,e^{i\,\omega\,t}+c.c \qquad  \theta(t)=\Theta\,e^{i\,\omega\,t}+c.c. \;\;\; ,\]
where $c.c.$ stands for complex conjugate.\\
Plugging the expressions above into Eqs.~(\ref{eq:y}) and  (\ref{eq:I})
we obtain an algebraic  linear system for the two 
(complex) amplitudes $Y$ and $\Theta$. The condition of having
nontrivial solutions provides 
a set of two real equations for the real and immaginary 
parts of $\omega$. Focusing
on the determination of marginal curves associated to flutter
conditions, and thus assuming the imaginary part of $\omega $ to be zero, 
one of the two equations furnishes the relationship between parameters 
identifying the marginal curve, e.g, $\rho$ as a function  of $k$ and $r$.\\
Due to the presence of Theodorsen's function the determination
of the marginal curve requires the exploitation of standard searching methods 
to find zeros of the associated (non polynomial) set of equations. This 
is however a task which can be easily achieved by means of a classical
Raphson--Newton method.

\begin{figure}
\centerline{\includegraphics[width=6cm]{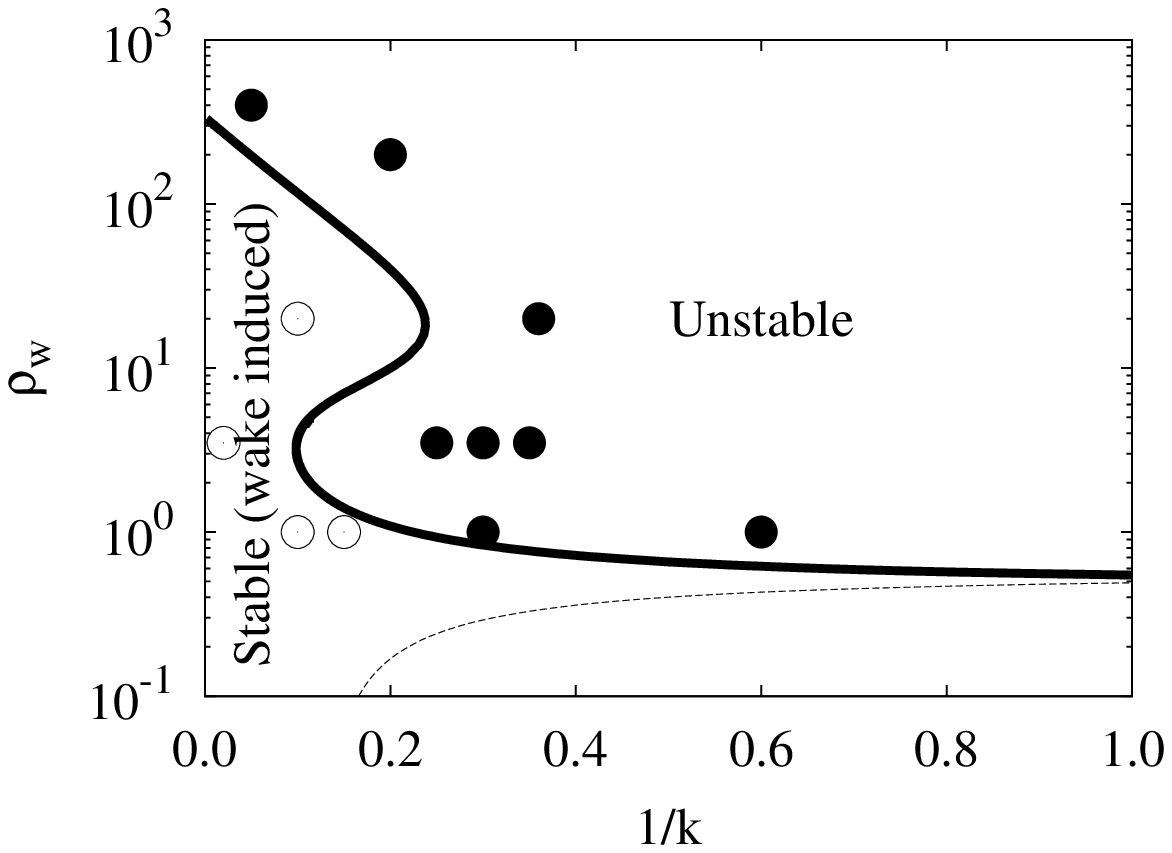}
\includegraphics[width=6cm]{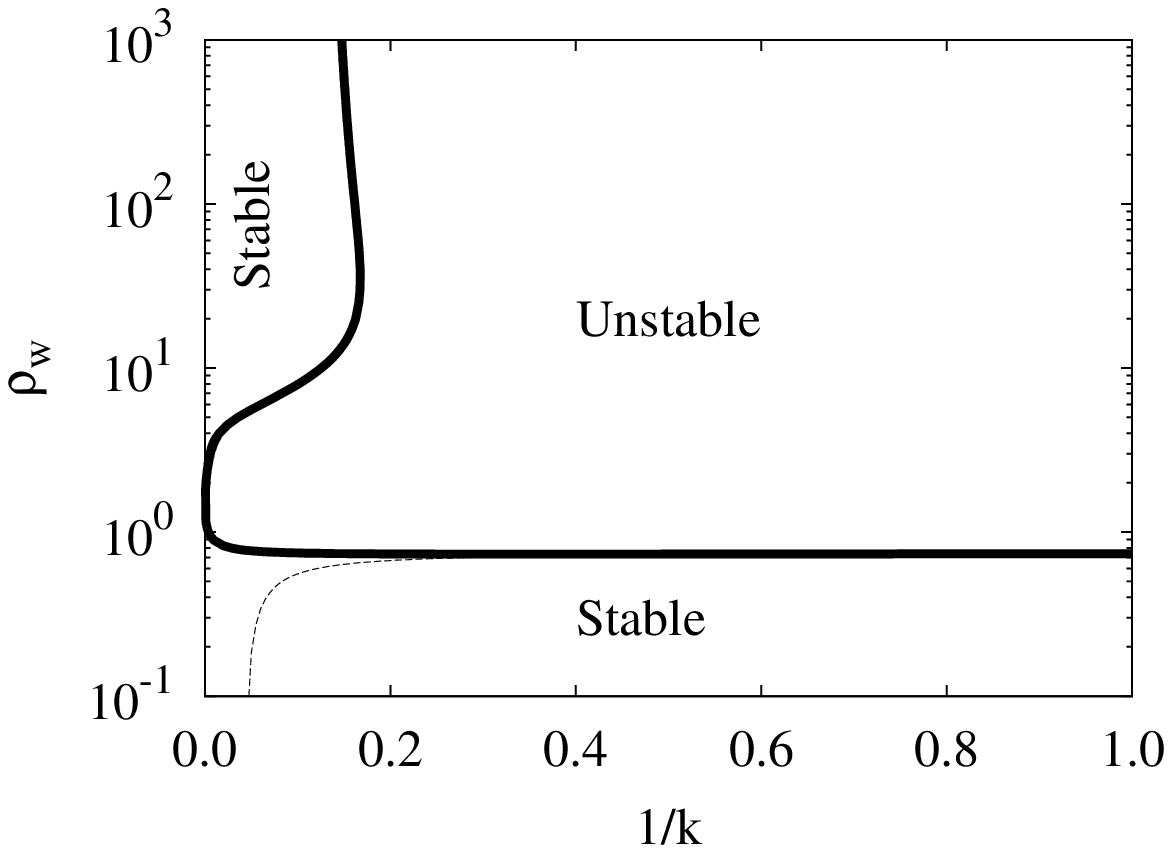}}
%\centerline{\includegraphics[width=6cm]{fig2a.eps}}
\caption{The marginal curve for $r=0.5$ (left, corresponding to the leading edge anchor point) and $r=0.26$ (right). Circles refer to points analyzed
through DNS. For filled circles DNS predict unstable motion; for
open circles they predict stability.  For comparison, 
dotted lines represent the marginal
curve under the quasi-steady hypothesis.
To emphasize variations along the vertical axis, a log scale is used
for  $\rho_w$ . 
}
%The segment represents the  transverse size $L_y$.}
\label{fig2}
\end{figure}

As customary, it is convenient to pass to a dimensionless form of all
variables. For this purpose, 
let us define the following dimensionless quantities: 
\begin{equation}
 t\,\frac{U}{L}\mapsto t\qquad \frac{y_C}{L}\mapsto y  \qquad \frac{k}{\rho\, U^2}\mapsto k \qquad  
\frac{\rho_w\,\delta}{\rho\, L}\mapsto \rho_w \;\;\; ,
\label{adimens}
\end{equation}
so that the reduced frequency is simply $q=\omega/2$.
To define the dimensionless parameters above, 
we have defined by $\rho$ the fluid density
and written the wing mass $m$ as $m=\rho_w \delta\,L$, $\rho_w$ 
being the wing density. \\
The behavior of the marginal curve for $r=0.5$ and $r=0.26$ is reported in
Fig.~\ref{fig2}. Circles refer to points analyzed
by DNS for Re=10000 and Re=60000. 
Their discussion is postponed to Sec.~\ref{numeric}.
Dotted lines represent the marginal
curves under the quasi-steady hypothesis (i.e. when ${\cal C}(q)=1$ and 
the equations defining marginal curves are polynomials).
This corresponds to neglecting the effect of the wake on
the motion of the wing.   Some points are worth discussing.
The first is on the shape of the marginal curves obtained. They are quite
structured and the relationship between the dimensionless 
parameters associated to the fluttering condition is
far from being trivial. The pronounced `nose' for $r=0.5$ 
around $\rho_w\sim 10$,  $1/k\sim 0.2$ is an example. This 
peculiar shape leads to interesting consequence on the fluttering
condition in terms, e.g.,  of the wing mass: there are indeed
three different values for the wing mass (actually for the dimensionless 
quantity $\rho_w\,\delta/(\rho\,L)$)
leading to fluttering for a fixed spring constant 
(actually for a fixed  dimensionless parameter $k/(\rho U^2)$).
It turns out that, for all spring constants and wind speed,  
it is impossible to sustain flapping
if  $\rho_w\,\delta/(\rho\,L)< 0.59$. The latter threshold has been determined by the
quasi-steady assumption that tends to coincide (see dotted line in 
Fig.~\ref{fig2}) with Theodorsen's  marginal curve
for small  spring constants. 
On the other hand, for very large
value of  $\rho_w\,\delta/(\rho\,L)$ (order of $300$ for $r=0.5$; a 
much larger value for  $r=0.26$) 
fluttering is expected at all velocities.\\
Interestingly, the conclusions above are very sensitive to the
value of $r$, i.e., to the relative distance between the wing center of mass 
and the anchor point (see right frame of Fig.~\ref{fig2}).

\begin{figure}
\centerline{\includegraphics[width=6cm]{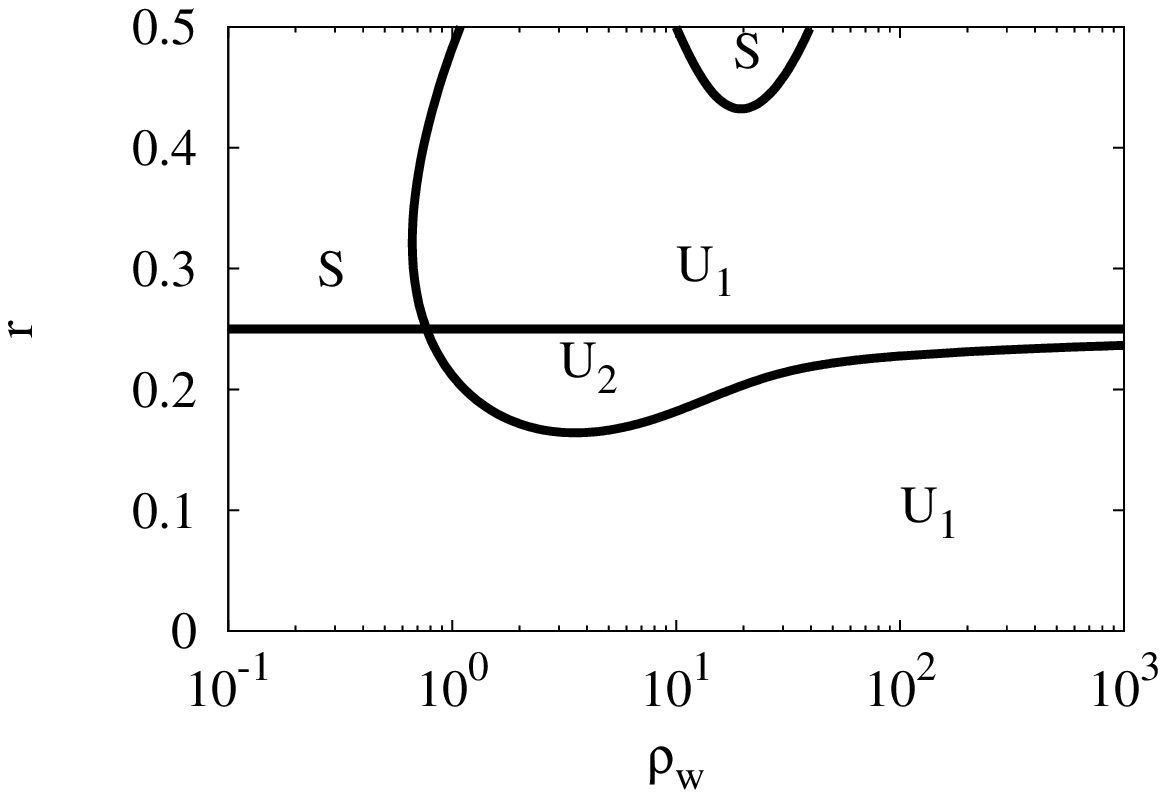}
\includegraphics[width=6cm]{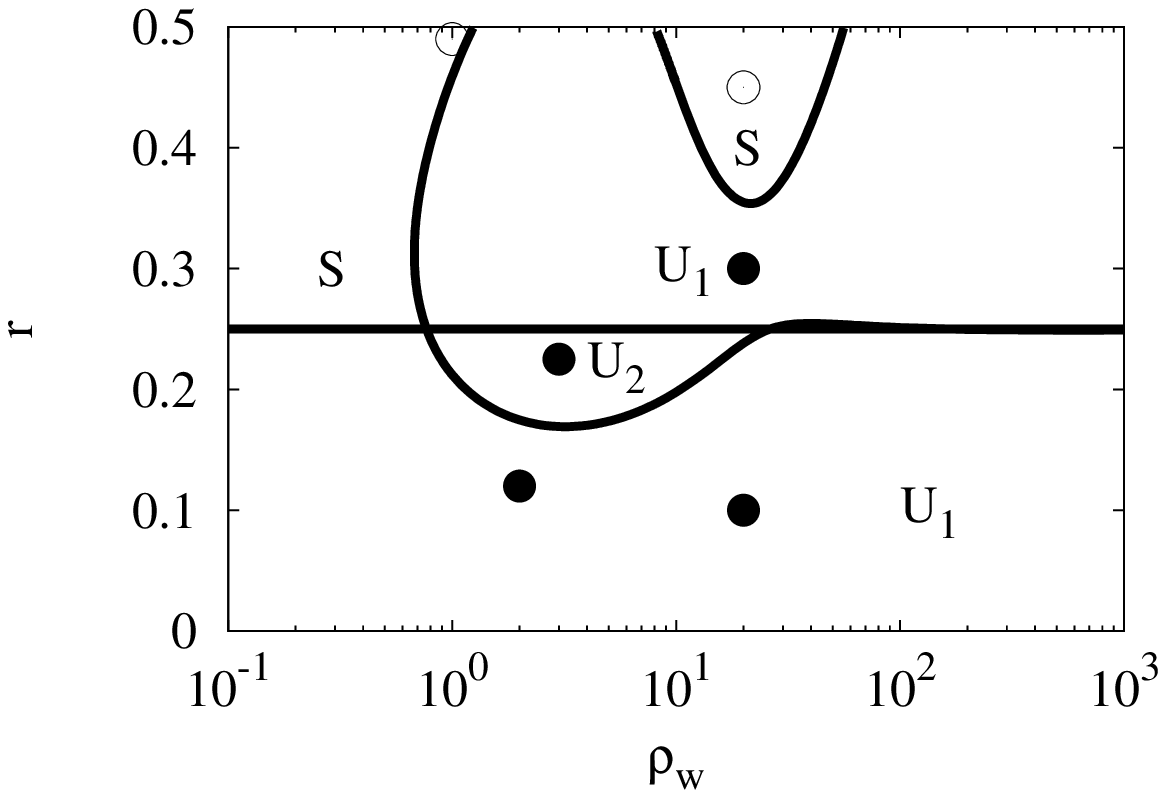}
}
%\centerline{\includegraphics[width=6cm]{fig2a.eps}}
\caption{The marginal curve for $k=1/0.2$ (left) and $k=1/0.17$ (right) 
in the $\rho_w$-$r$ space.
`$U_1$' and `$U_2$' denote Unstable regions with one and two unstable normal modes, respectively;
`S' stands for Stable. 
For filled circles DNS predict unstable motion; for
open circles they predict stability. 
%The dotted line
%is the line corresponding to the zero frequency regime
%occurring at $r=0.25$.
}
%The segment represents the  transverse size $L_y$.}
\label{fig3}
\end{figure}

To better understand the sensitivity to $r$ we report in 
Fig.~\ref{fig3} the marginal curves in the plane $\rho_w$-$r$
for two fixed values of $k$. The non-trivial role played
by the couple $\rho_w$-$r$ in the emergence of stable/unstable
regimes can be clearly detected from the figure.

\section{Direct numerical simulations}
\label{dns}

The aim of this section is to give a brief description of the 
numerical method used to numerically approximate the laminar
incompressible Navier--Stokes equations 
(so in essence we are doing direct numerical simulations or DNS) and 
verify the predictions obtained from the normal mode linear 
stability analysis with the solutions obtained from the DNS simulations.\\
It is beyond the scope of the present paper
to fully analyze the shape and position of the obtained marginal
curves in the whole parameter space  $k$-$\rho_w$. This issue is indeed 
quite expensive from the numerical viewpoint and it will be eventually 
postponed to a possible future analysis.
Rather, our aim is to corroborate our former, and to some extent surprising,
conclusions on  the nontrivial shape of the marginal curves.

\subsection{The numerical method}
Hereupon, we briefly outline the solution methodology used to 
solve the governing equations on moving overlapping structured grids. 
The complete description of the numerical method and gridding methodology 
can be found in the papers by \citet{Henshaw1994} and \citet{Chesshire1990}.\\
In this manuscript, we numerically approximate the laminar incompressible 
Navier--Stokes equations by using
the velocity-pressure formulation or pressure-Poisson equation (PPE) 
\citep{Henshaw1994,Sani2006,Petersson2001,Gresho1991}.
\begin{eqnarray}
\frac{\partial \textbf{u} }{\partial t} + \textbf{u} 
\cdot \bm{\partial} \textbf{u}  = -\frac{\bm{\partial} p}{\rho} + \nu 
\partial^{2} 
\textbf{u}
\quad &\textit{for} \quad \textbf{x} 
\in \mathcal{D}, \quad t>0,   \label{equ2_5}  \\ 
{\frac{\partial^{2} p}{\rho}} 
+ \partial_{\alpha} \textbf{u}\cdot \bm{\partial} u_{\alpha}=  0  
\quad &\textit{for} \quad \textbf{x} \in 
\mathcal{D}, \quad t>0,  
\label{equ2_6} 
\end{eqnarray}
\noindent with the following boundary and initial conditions
\begin{eqnarray}B 
\left(  \textbf{u}, p \right) &= 0  \qquad\qquad\qquad  
&\textit{for} \quad \textbf{x} \in \partial\mathcal{D}, 
\quad\,\,\, t>0, 
\label{equ2_7}  \\\bm{\partial} \cdot \textbf{u} &= 0  
\qquad\qquad\qquad &\textit{for} \quad 
\textbf{x} \in \partial\mathcal{D}, \quad\,\,\, t>0, 
\label{equ2_8}  \\\textbf{u}
\left( \textbf{x},0 \right)  &= \textbf{u}_0 \left( \textbf{x} \right) 
\qquad\qquad\quad\,\,  &\textit{for} \quad \textbf{x} \in \mathcal{D} .  
\label{equ2_9}
\end{eqnarray}
In this initial-boundary-value problem (IBVP), $\textbf{x}=(x,y)$ (for $\mathbb{N}=2$ where 
$\mathbb{N}$ is the number of space dimensions) is the physical space 
coordinates, 
%$\mathcal {P}=\mathcal {P}(x,y,z,t)$
$\mathcal{D}$ is a bounded domain in $\Re^{\mathbb{N}}$, 
$\partial\mathcal{D}$ is the boundary of the domain $\mathcal{D}$, $t$ is 
the physical time, $\textbf{u}=(u,v)$ is the velocity field, 
$p$ is the pressure, $\nu$ is the kinematic viscosity, 
$\rho$ is the density, $B$ is a boundary operator
%$, \textbf{g}$ is the  boundary data and $\textbf{q}_0$ is the 
initial data.\\
and $\textbf{u}_0$ are the initial conditions. Hence, we look for an approximate numerical 
solution of equations (\ref{equ2_5}) and (\ref{equ2_6}) in a given 
domain $\mathcal{D}$,   with prescribed boundary conditions and 
given initial conditions  (equations   (\ref{equ2_7})-(\ref{equ2_9})).  
Equations (\ref{equ2_5})-(\ref{equ2_9}) are solved in logically 
rectangular grids in the transformed computational space  
$\mathcal {C}=\mathcal {C}(\xi,\eta,t)$ (refer to 
\citet{Chesshire1990,Drikakis2004,Guerrero2009,Guerrero2010,Henshaw1994,Vinokur1974}
for a detailed derivation), using second-order 
centred finite-difference approximations on structured 
overlapping grids.  \\
In general, the motion of the component grids $\mathcal{G}_g$ of an
overlapping grid system  $\mathbb{G}= \{\mathcal{G}_g\} $, may be an 
user-defined time dependent function, may obey the Newton-Euler equations for 
the case of rigid body motion or may correspond to the boundary nodes displacement 
in response to the forces exerted by the fluid pressure for the case of fluid-structure 
interaction problems.  For moving overlapping grids, Eqs.~ 
(\ref{equ2_5})-(\ref{equ2_6}) are expressed in a reference frame moving 
with the component grid   as follows,
\begin{eqnarray}
\frac{\partial \textbf{u} }{\partial t} + \left[ \left( \textbf{u} - 
\dot{\textbf{G}} 
\right)   \cdot \bm{\partial}  \right] \textbf{u}  =  
-\frac{\bm{\partial} p}{\rho} + 
\nu \partial^{2} \textbf{u} \quad & \mbox{for} \quad \textbf{x} \in 
\mathcal{D}, \quad  t>0,   \label{equ2_11}  \\ 
{\frac{\partial^{2} p}{\rho}} 
+ \partial_{\alpha} \textbf{u}\cdot \bm{\partial} u_{\alpha}=  0  
 \quad &\mbox{for} \quad \textbf{x} 
\in \mathcal{D}, \quad t>0,  
\label{equ2_12} 
\end{eqnarray}
\noindent where $\dot{\textbf{G}}$ is the rate of change of the position 
of a given set of grid points $\textbf{x}_{p}^{g}$ of a component grid $\mathcal{G}_g$   in the physical 
space (grid velocity).  It is important to mention that 
the new governing equations expressed in the moving reference frame must 
be accompanied by proper boundary conditions. For a moving body with a 
corresponding moving no-slip wall, only one constraint may be applied 
and this corresponds to the velocity on the wall, such as
\begin{equation}\textbf{u} \left( \textbf{x}_p^{g} \vert_{wall}, t\right)    =    
\dot{\textbf{G}} \left( \textbf{x}_p^{g} \vert_{wall}, t  \right) , \quad \mbox{where} \quad   
\textbf{x}_p^{g}  \vert_{wall}  \in \partial \mathcal{D}_{wall} \left( t \right). 
\label{equ2_13}
\end{equation}
Finally, in order to keep the solution of the pressure equation decoupled 
from the solution of the velocity components, we choose a time stepping 
scheme for the velocity components that only involves the pressure from 
the previous time steps (split-step scheme) \citep{Henshaw1994,Henshaw2003b}.  
Then, the discretized equations 
are integrated in time using a semi-implicit multi-step method, that uses a 
Crank-Nicolson scheme for the viscous terms and a second-order 
Adams-Bashforth/Adams-Moulton predictor-corrector approach for 
the convective terms and pressure.  This solution method yields 
a stable second-order accurate in space and time numerical scheme on moving 
overlapping structured grids. \\
To assemble the overlapping grid system 
$\mathbb{G}$  and solve the laminar incompressible Navier-Stokes 
equations in their velocity-pressure formulation, the 
Overture\footnote{https://computation.llnl.gov/casc/Overture/} 
framework is used.  The large sparse non-linear system of equations 
arising from the discretization of the laminar incompressible 
Navier-Stokes equations is solved using the 
PETSc\footnote{http://www-unix.mcs.anl.gov/petsc/petsc-as/} 
library, which was interfaced with Overture.  The system of 
non-linear equations is then solved using a Newton-Krylov 
iterative method, in combination with a suitable preconditioner.

As a final remark, for the range of Reynolds numbers considered 
in this study, no turbulence model or sub-grid scale model is used 
for the numerical simulations conducted, 
so in essence two-dimensional direct-numerical simulations (DNS) are 
performed.  Despite the fact that the grid does not resolve the smallest scales
in the far-wake region 
(where small scale activity is however strongly reduced), 
the grid resolution is always more 
than adequate. It is so also in the near-wake flow field
where the resolution is sufficiently high to properly resolve 
the small scales 
so that accurate instantaneous drag and lift forces are computed.

\subsection{Numerical settings and results}
\label{numeric}
We confine our attention to the marginal curve of 
Fig.~\ref{fig2} (left), performing
direct numerical simulations for different
values of $(\rho_w$, $1/k)$, for Reynolds numbers between 
$10000$ and $60000$. We almost never place ourselves
too close to the predicted marginal line to avoid a very 
slow increase/decrease
of the initial perturbation and thus to avoid 
long and expensive simulations. As far as initial perturbations are 
concerned, 
we  firstly consider the one dimensional problem (along $x$)
and we reach the equilibrium position under that condition. 
At this stage we impose a perturbation on 
the angle  $\theta$, of the order of $10^{-5}\, deg$, and 
leave the system free to evolve. \\
\begin{figure}
\centerline{\includegraphics[width=6cm]{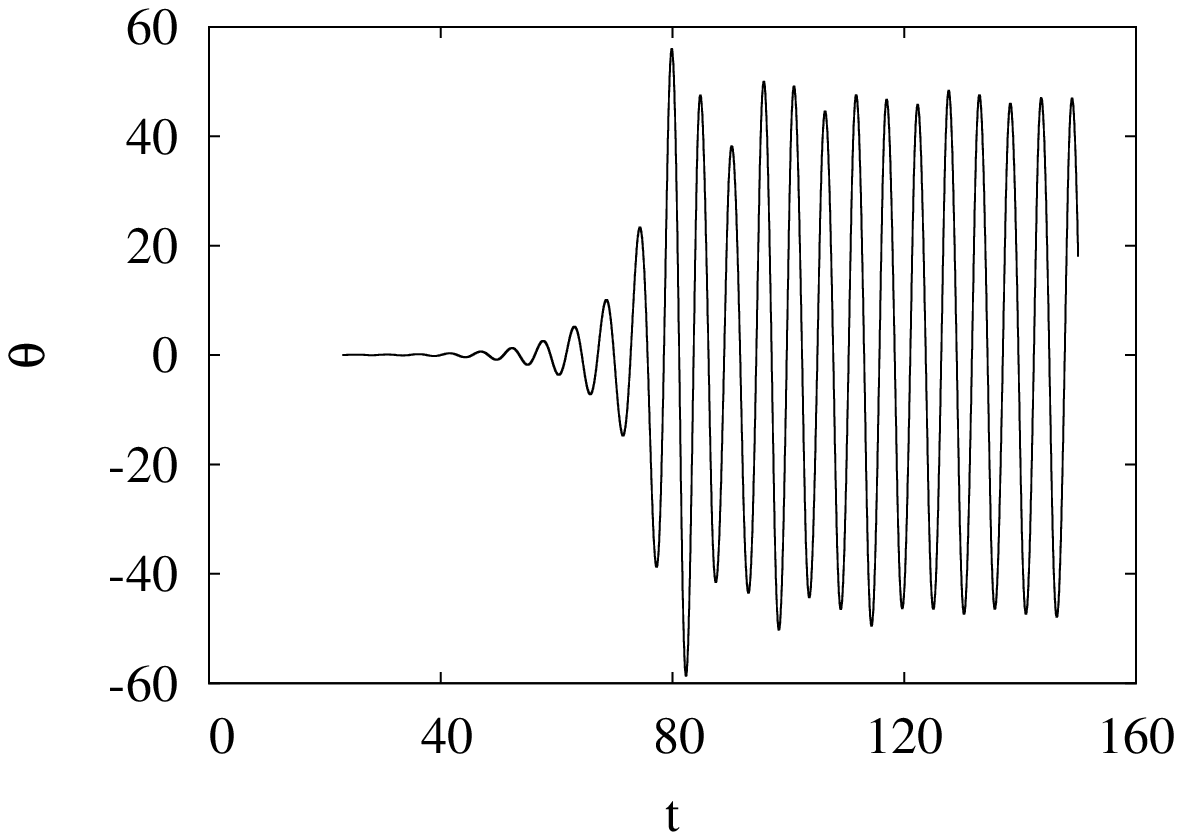}
\includegraphics[width=6cm]{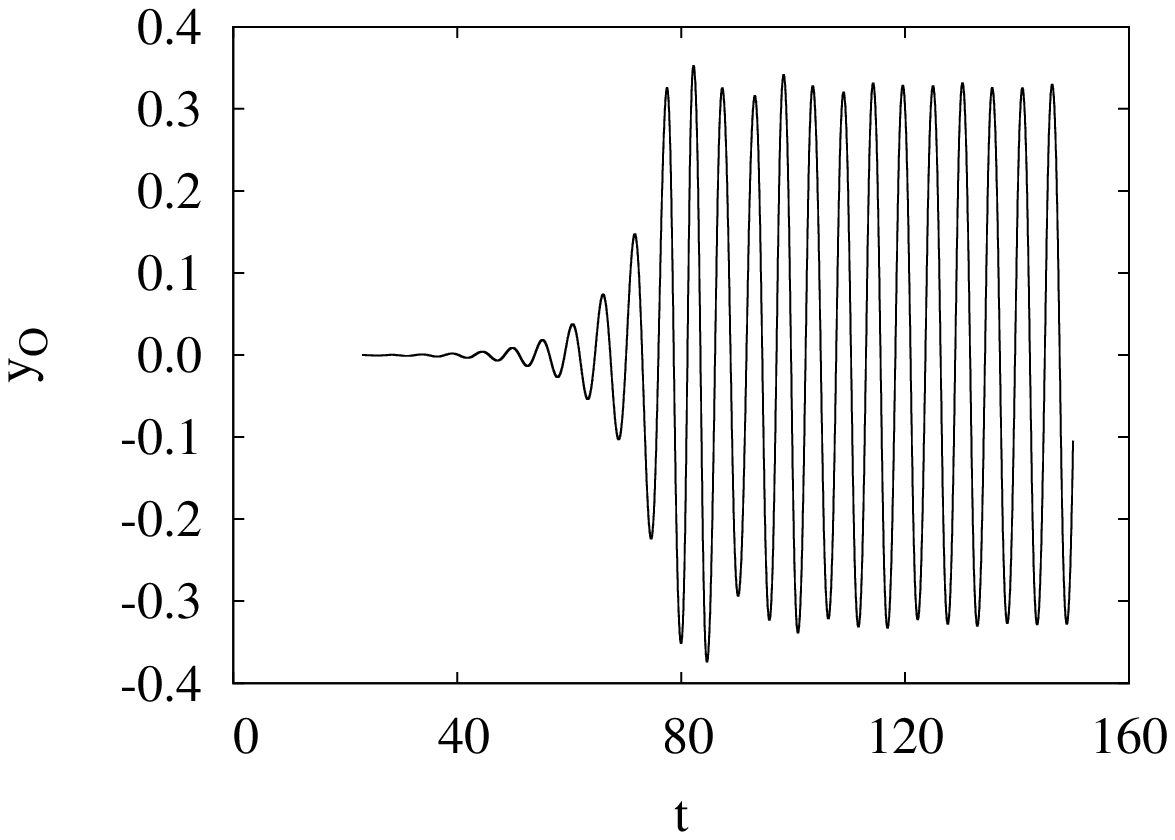}
}
\caption{Time behavior of the pitching angle $\theta$ (deg), on the left, 
and of the $y$-coordinate of the wing leading edge, on the right, for the
unstable case with $1/k=0.35$ and $\rho_w=3.5$ and $r=1/2$. 
All variables are made  dimensionless by (\ref{adimens}). 
}
\label{fig4}
\end{figure}
\begin{figure}
\centerline{\includegraphics[width=6cm]{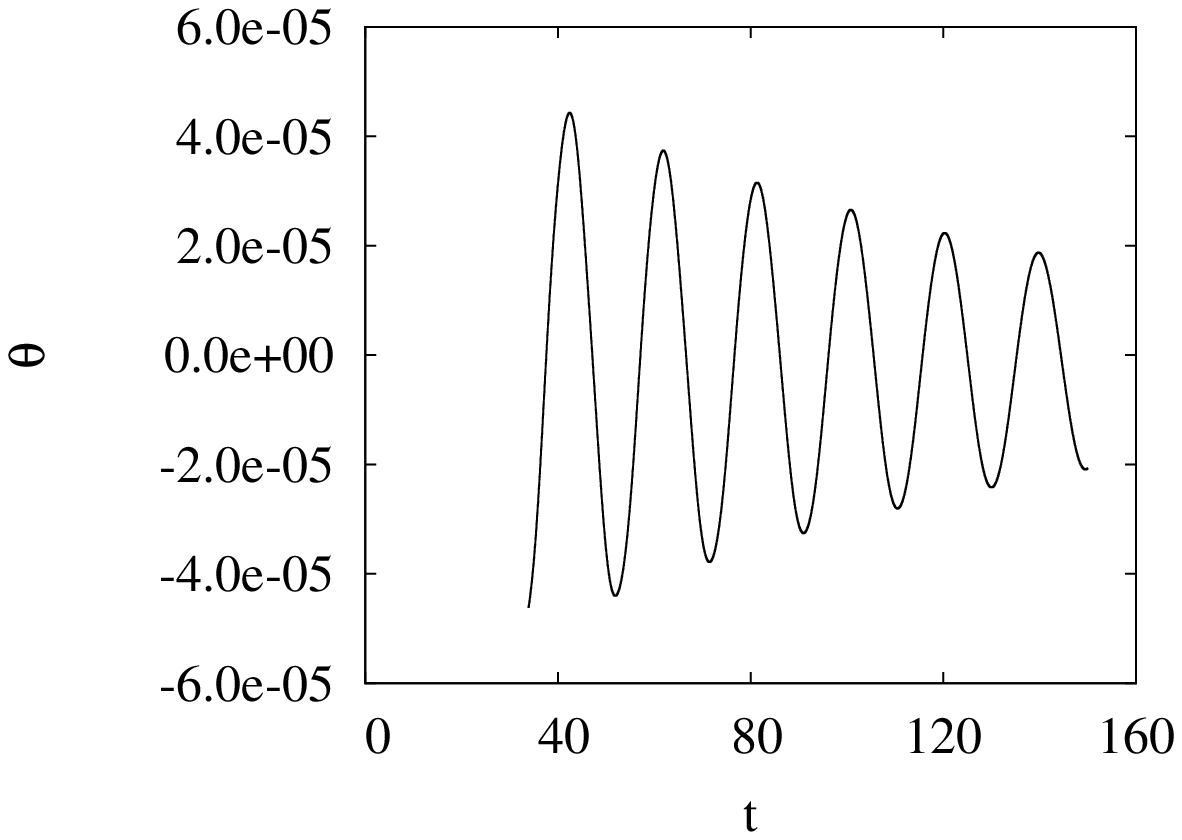}
\includegraphics[width=6cm]{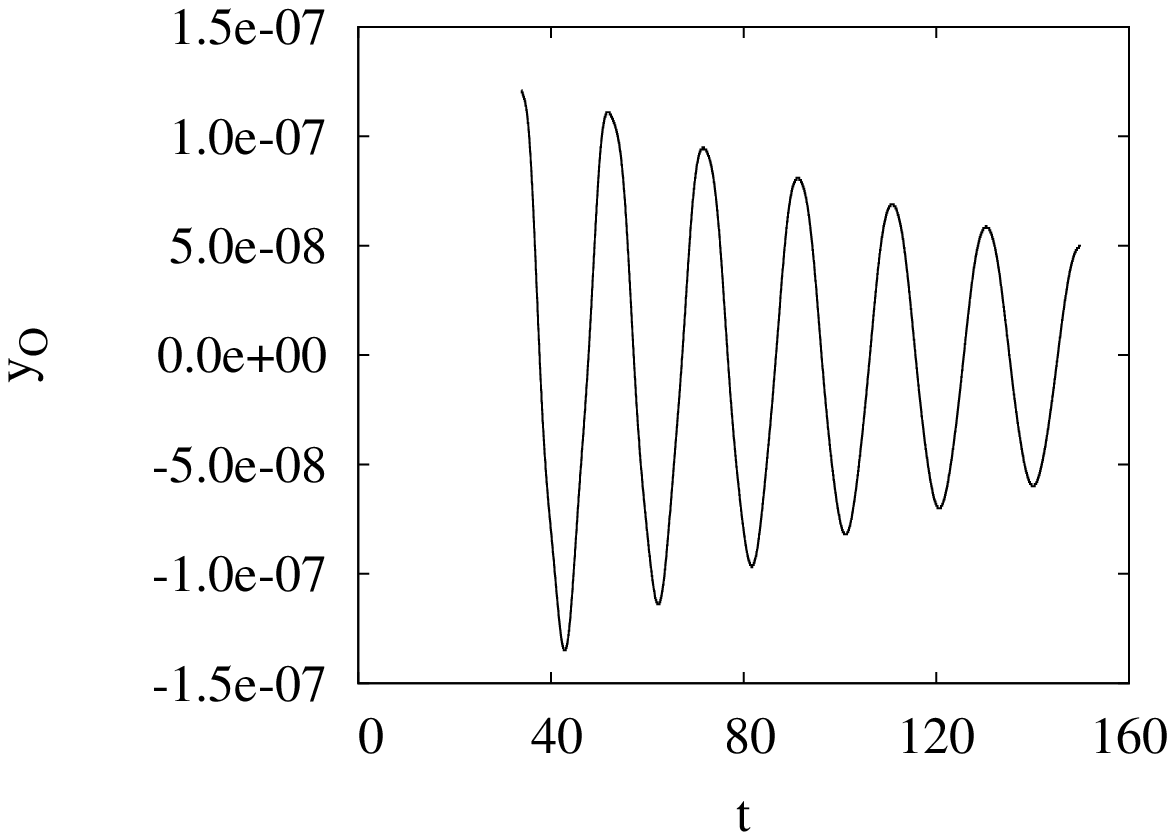}
}
\caption{The same as in Fig.~\ref{fig4} but for the stable case 
 $1/k=0.1$ and $\rho_w=20$. 
}
\label{fig5}
\end{figure}
Filled circles in  Fig.~\ref{fig2} (left)
are associated to unstable behavior; stability is represented 
by  open circles.  The agreement with our predictions based on 
Theodorsen's theory is good.\\
Typical time behavior for the angle $\theta$ and the leading edge
 vertical coordinate $y_O$
are reported in Fig.~\ref{fig4}  for $\rho_w=3.5$ and $1/k=0.35$
(unstable case, see also the movie) 
and in Fig.~\ref{fig5}  for $\rho_w=20$ and $1/k=0.1$
(stable case, see also the movie). 
Notice how the unstable case reaches a fully nonlinear stage
characterized by a sustained flapping with a $y$ excursion of the wing 
leading edge of order one (with lengths normalized by the
wing chord). This is a very promising result in view of applications
related to the energy harvesting where elastomeric capacitors need to be
stretched/compressed much, in order to produce reasonable amounts of 
electric energy.\\
The example shown in Fig.~\ref{fig4} tells us that an infinitesimal
perturbation yields the system to a finite-amplitude limit cycle
associated to unceasing flapping.
A natural question which arises is on whether  
unceasing flapping can be observed
if a finite size perturbation is applied to configurations  which are stable
with respect to small perturbations. As an example we address this question
for the couple of parameters 
$\rho_w=20$ and $1/k=0.1$.  The initial angle $\theta$ is now
$90\ deg$. The resulting time behaviors for $\theta$ and $y_O$ are reported in
Fig.~\ref{fig6}. We do not detect  any tendency of relaxation toward the
aligned configuration, a fingerprint of a subcritical instability.
\begin{figure}
\centerline{\includegraphics[width=6cm]{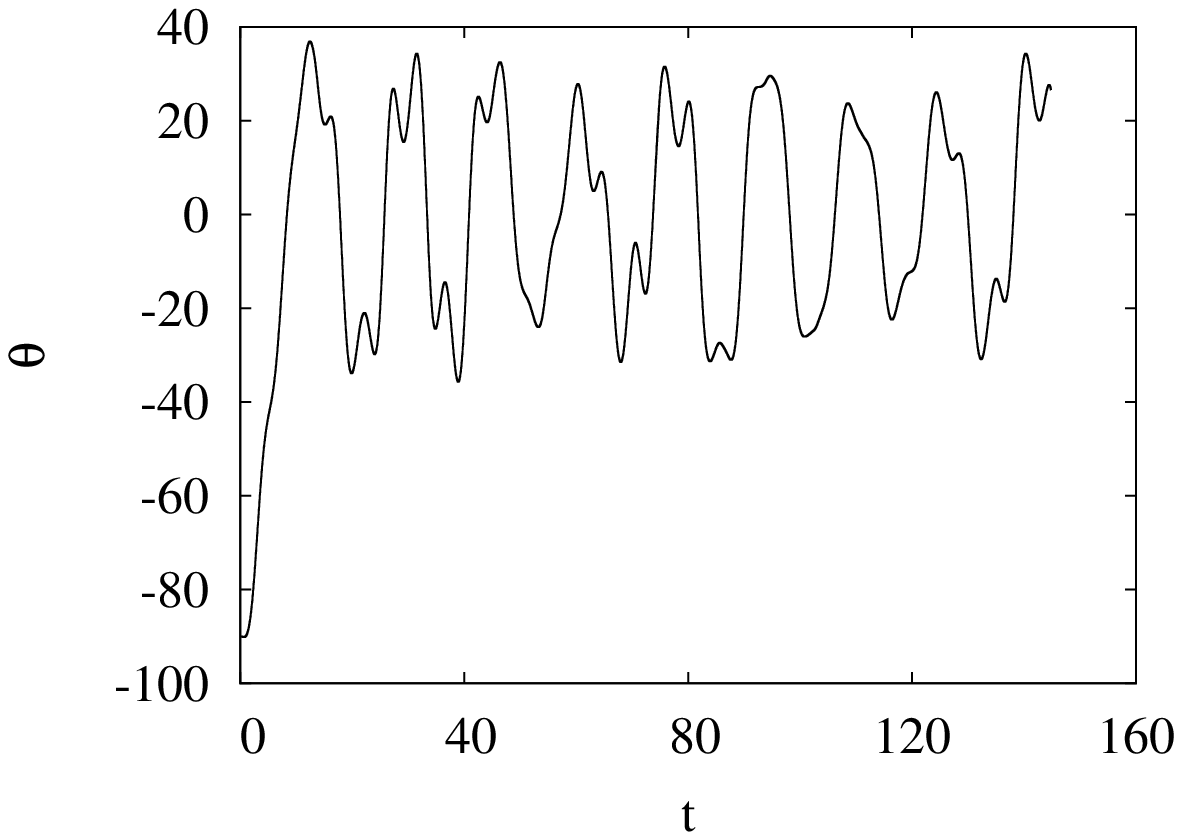}
\includegraphics[width=6cm]{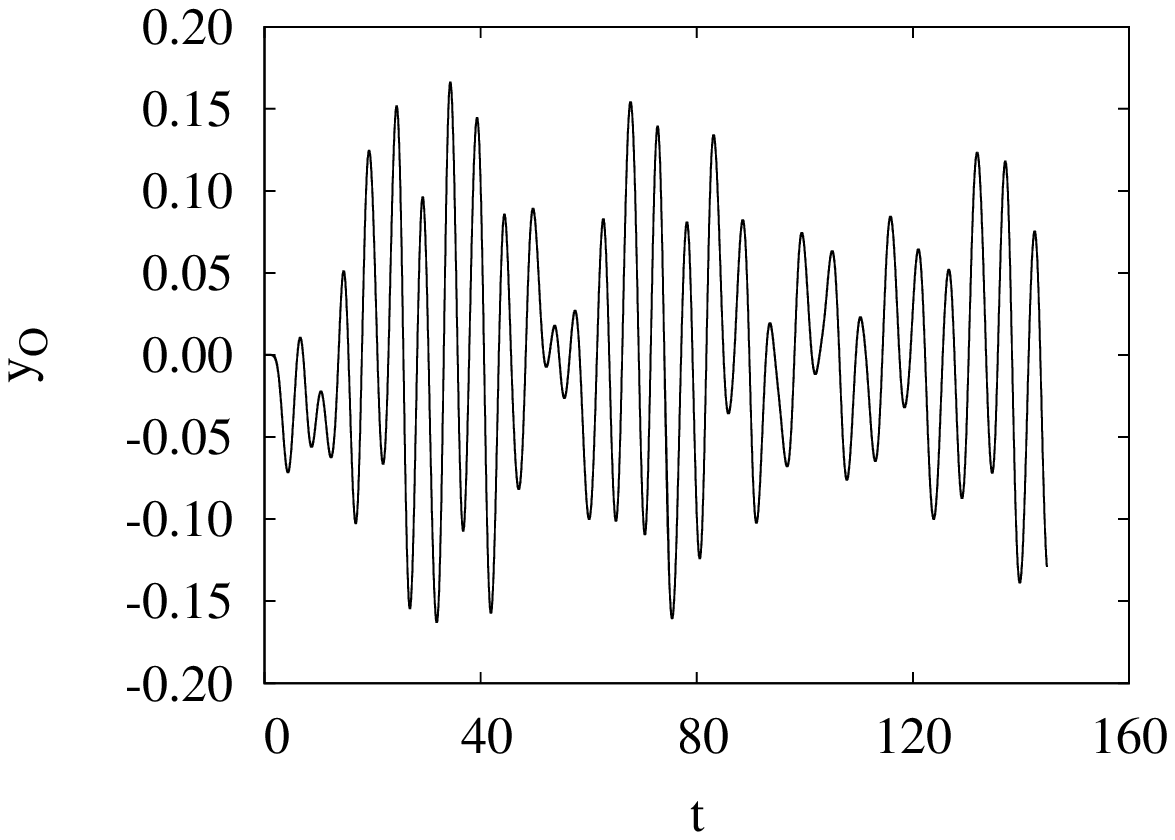}
}
\caption{Time behavior of the pitching angle $\theta$ (deg), on the left, 
and of the $y$-coordinate of the wing leading edge, on the right, for the
linearly stable case with $1/k=0.1$, $\rho_w=20$ and $r=1/2$. 
All variables are made 
dimensionless by (\ref{adimens}). The initial perturbation is here of finite size. 
}
\label{fig6}
\end{figure}

\section{Conclusions}
A simple flapping system has been presented and investigated both analytically
and numerically.
Three dimensionless parameters come into play and their role on the resulting
flapping states appear to be  highly non trivial.  This 
property has been confirmed by numerical simulations 
we carried out in the range of Reynolds 
numbers between $10000$ and $60000$.\\
Both supercritical instabilities and subcritical instabilities 
have been  found  in our system. 

Our findings have direct applicability to 
the energy harvesting problem by fluid-structure interaction \citep{Borex012}.
We have found sets of parameters leading to unceasing flapping for which the 
excursion  of the leading edge is of the same order of the wing chord.
Once our spring is replaced by an elastomeric capacitor, the observed
large amplitude of the wing oscillations is a necessary condition for
extracting  a reasonable 
amount of energy from the  device.

\vspace{0.8cm}
%\noindent {\it Acknowledgements}\\
We thank Alessandro Bottaro and Jan Oscar Pralits 
for many useful discussions and suggestions.
The use of the computing resources at CASPUR high-performance computing center
was possible thanks to the HPC Grant 2011. The use of the computing facilities
at the high-performance computing center of the University of Stuttgart was
possible thanks to the support of the HPC-Europa2 project (project number
238398), with the support of the European Community – Research Infrastructure
Action of the FP7.

%%%%%%%%%%%%%%%%%%%%%%%%%%%%%%%%%%%%%%%%%%%%%%%%%%%%%%%%%%%%%%%%
\bibliographystyle{jfm}
\bibliography{biblio}

\end{document}